\documentclass[sigconf, authorversion]{acmart}

\AtBeginDocument{%
  \providecommand\BibTeX{{%
    \normalfont B\kern-0.5em{\scshape i\kern-0.25em b}\kern-0.8em\TeX}}}

\usepackage{todonotes}
\usepackage[nolist,nohyperlinks ]{acronym}
\usepackage{siunitx}
\usepackage{calculator}
\usepackage{csquotes}
\usepackage{subcaption} 
\usepackage{fmtcount}
\usepackage{enumitem}
\setlist[description]{%
labelindent=0.5\parindent,%
itemindent=-.6em,%
leftmargin=*,%
}

\begin{acronym}[UMLX]
	\acro{HMD}{head-mounted display}
	\acro{GUI}{graphical user interface}
 	\acro{UI}{user interface}
	\acro{HUD}{head-up display}
	\acro{TCT}{task-completion time}
	\acro{SVM}{support vector machine}
	\acro{EMM}{estimated marginal mean}
	\acro{AR}{Augmented Reality}
 	\acro{MR}{Mixed Reality}
	\acro{VR}{Virtual Reality}
    \acro{VE}{Virtual Environment}
	\acro{CSCW}{Computer-Supported Cooperative Work}
	\acro{HCI}{Human-Computer Interaction}
	\acro{ABI}{Around-Body Interaction}
	\acro{SLAM}{Simultaneous Location and Mapping}
	\acro{ROM}{range of motion}
	\acro{EMG}{electromyography}
	\acro{STS}{System Trust Scale}
	\acro{IPQ}{Igroup Presence Questionnaire}
	\acro{TLX}{NASA Task Load Index}
 	\acro{RTLX}{Raw Nasa-TLX}
 	\acro{CAD}{Computer Aided Design}
    \acro{ART}{Aligned Rank Transform}
	\acro{IOS}{Inclusion of Other in the Self}
	\acro{GEQ}{Game Experience Questionnaire}
\end{acronym}

\copyrightyear{2024}
\acmYear{2024}
\setcopyright{acmlicensed}\acmConference[CHI '24]{Proceedings of the CHI Conference on Human Factors in Computing Systems}{May 11--16, 2024}{Honolulu, HI, USA}
\acmBooktitle{Proceedings of the CHI Conference on Human Factors in Computing Systems (CHI '24), May 11--16, 2024, Honolulu, HI, USA}
\acmDOI{10.1145/3613904.3642864}
\acmISBN{979-8-4007-0330-0/24/05}
\newcommand{\systemname}{Just Undo It}
\newcommand{\longtitle}{Exploring Undo Mechanics in Multi-User Virtual Reality}

\newcommand{\ivUndoTechnique}{\textsc{UndoTechnique}}

\newcommand{\ivCollaborationMode}{\textsc{CollaborationMode}}

\newcommand{\ivCollaborative}{\textsc{Collaborative}}
\newcommand{\ivDivided}{\textsc{Divided}}

\newcommand{\ivNoUndo}{\textsc{NoUndo}}
\newcommand{\ivIndividualUndo}{\textsc{IndividualUndo}}
\newcommand{\ivSelectiveUndo}{\textsc{SelectiveUndo}}
\newcommand{\ivWorldUndo}{\textsc{WorldUndo}}

\newcommand{\dvIOS}{\textsc{\acs{IOS}}}
\newcommand{\dvGEQ}{\textsc{\acs{GEQ}}}
\newcommand{\dvTLX}{\textsc{\acs{RTLX}}}

\newcommand{\dvDisturbing}{\textsc{DisturbingOthers}}
\newcommand{\dvDisturbed}{\textsc{FeltDisturbed}}
\newcommand{\dvControl}{\textsc{Control}}
\newcommand{\dvFrustration}{\textsc{Frustration}}
\newcommand{\dvSuccess}{\textsc{Success}}
\newcommand{\dvRecover}{\textsc{Recover}}

\newcommand{\dvEnjoyment}{\textsc{Enjoyment}}
\newcommand{\dvDesiredFutureUsage}{\textsc{DesiredFutureUsage}}

\newcommand{\dvTowerHeight}{\textsc{TowerHeight}}
\newcommand{\dvNumberUndos}{\textsc{NumberOfUndos}}

\newcommand{\dvNumberGrabs}{\textsc{NumberOfGrabs}}

\newcommand{\ano}[4]{$F_{#1, #2}=#3$, $p#4$}

\newcommand{\subEtaG}[2]{%
	\ifthenelse{\equal{#1}{\string >.05}}
	{}
	{, $\eta_{G}^{2}=#2$}%
}

\newcommand{\subEta}[2]{%
	\ifthenelse{\equal{#1}{\string >.05}}
	{}
	{, $\eta^{2}=#2$}%
}

\newcommand{\chisq}[3]{$\chi^2(#1) = #2$, $p #3$}

\def\ges{$\eta_{G}^{2}$}

\newcommand{\efETAsquared}[1]{%
	\ifdim#1pt>0.139pt 
	large (\ges{} = #1)
	\else 
	\ifdim#1pt>0.059pt 
	medium (\ges{} = #1)
	\else 
	small (\ges{} = #1)
	\fi
	\fi
}
\usepackage{etoolbox}

\newcommand{\nquote}[1]{\enquote{\emph{#1}}}

\begin{document}

\title[\systemname{}]{\systemname{}: \longtitle{}}

\settopmatter{authorsperrow=4}

\author{Julian Rasch}
\orcid{0000-0002-9981-6952}
\affiliation{
  \institution{LMU Munich}
  \streetaddress{Frauenlobstr. 7A}
  \city{Munich}
  \country{Germany}
  \postcode{80337}
}
\email{julian.rasch@ifi.lmu.de}

\author{Florian Perzl}
\orcid{0009-0006-6878-0949}
\affiliation{
  \institution{LMU Munich}
  \streetaddress{Frauenlobstr. 7A}
  \city{Munich}
  \country{Germany}
  \postcode{80337}
}
\email{florian.perzl@campus.lmu.de}

\author{Yannick Weiss}
\orcid{0000-0002-7332-3287}
\affiliation{
 \institution{LMU Munich}
  \streetaddress{Frauenlobstr. 7A}
  \city{Munich}
  \country{Germany}
  \postcode{80337}
}
\email{yannick.weiss@ifi.lmu.de}

\author{Florian Müller}
\orcid{0000-0002-9621-6214}
\affiliation{
  \institution{LMU Munich}
  \streetaddress{Frauenlobstr. 7A}
  \city{Munich}
  \country{Germany}
  \postcode{80337}
}
\email{florian.mueller@ifi.lmu.de}

\renewcommand{\shortauthors}{Rasch et al.}


\begin{abstract}
With the proliferation of VR and a metaverse on the horizon, many multi-user activities are migrating to the VR world, calling for effective collaboration support. As one key feature, traditional collaborative systems provide users with undo mechanics to reverse errors and other unwanted changes. While undo has been extensively researched in this domain and is now considered industry standard, it is strikingly absent for VR systems in research and industry.
This work addresses this research gap by exploring different undo techniques for basic object manipulation in different collaboration modes in VR. We conducted a study involving 32 participants organized in teams of two. Here, we studied users' performance and preferences in a tower stacking task, varying the available undo techniques and their mode of collaboration. The results suggest that users desire and use undo in VR and that the choice of the undo technique impacts users' performance and social connection.



\end{abstract}


\begin{teaserfigure}
	\includegraphics[width=\textwidth]{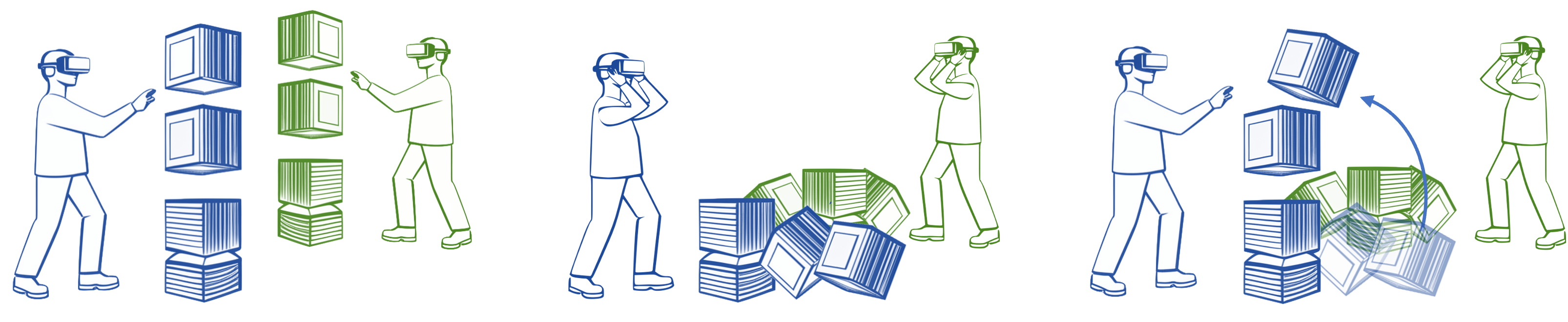}
	\caption{In this paper we explore the influence of different implementations of undo functionalities for multi-user Virtual Reality in varying modes of collaboration on the users' effectiveness, efficiency and social connection.}
	\Description[Teaser Figure for Just Undo It]{This three part teaser figure of the paper consists of three parts. The first part shows a green and a blue person constructing a tower each. In the second part both towers collapsed and both persons are in despair about this. In the third part the blue tower reverts back to its state before the collapse, the green tower however remains collapsed and the green person unhappy about it.}
	\label{fig:teaser}
\end{teaserfigure}

\maketitle

\section{Introduction}
\label{sec:introduction}







With increasing numbers of social \ac{VR} applications, both in the personal entertainment domain \cite{wienrich_social_2018} as well as the professional training and qualification domain \cite{de_paolis_multi-user_2018, schild_applying_2018}, \ac{CSCW} in \ac{VR} increases as well. In these shared \ac{VR} environments users can co-exist or collaborate with others. As a consequence, individual as well as collaborative tasks intertwine in many scenarios in the shared \ac{VR} and can affect other users of the environment. Like in the physical world, small errors caused by one user or the technical system (e.g. Tracking Errors) can destroy the progress of others and lead to undesired changes in the \ac{VR} environment.



In traditional computer systems, one established solution to this challenge are undo actions \cite{archer_user_1984, teitelman_automated_1972}, reverting the system or parts of the system to a previous state. While originally this is a single-user feature, subsequent work from \ac{CSCW} extends this \textit{personal or individual undo} to be suitable for collaborative work as well, by extending the range of effect beyond the individual actions. 
While many other established mechanics from traditional \ac{UI}s, like pointing \cite{pfeuffer_gaze_2017}, selecting \cite{argelaguet_survey_2013}, and manipulating \cite{poupyrev_egocentric_1998}, are well researched and established in collaborative \ac{VR} \cite{pinho_cooperative_2008, laviola_3d_2017}, the usage of undo functions remains niche in \ac{VR} applications. Besides only a few recent exceptions \cite{muller_undoport_2023, zhang_vrgit_2023} addressing specific use cases, undo features in \ac{VR} are not investigated by researchers yet.




In this paper, we revisit questions from \ac{CSCW} to identify how to design undo mechanics for multi-user \ac{VR} and how users will utilize them.
Based on a thorough review of related work we compare three undo techniques for \ac{VR} object manipulation, namely \ivIndividualUndo{}, \ivSelectiveUndo{}, \ivWorldUndo{} and a \ivNoUndo{} baseline, each with different effect range in two different modes of collaboration.
In our user study (n=32) 16 pairs of participants work \ivDivided{} and \ivCollaborative{} on a tower construction task with one \ivUndoTechnique{} or the \ivNoUndo{} baseline available. Our aim is to determine users’ approval of the proposed undo concepts in the different \ivCollaborationMode{}s and their willingness to use these features in the future. Additionally, we seek to understand how the user experience of the \ivUndoTechnique{}s changes based on the type of cooperation, considering factors such as efficiency, recoverability, users' willingness to take risks, intuitiveness, user expectations, and enjoyment.

The contribution of our paper is two-fold. First, we contribute the results of a controlled experiment exploring the strengths and weaknesses of the proposed \ivUndoTechnique{}s for different \ivCollaborationMode{}s. Second, based on our results, we derive design recommendations for the implementation of multi-user \ivUndoTechnique{}s in \ac{VR}.
Our results indicate, that like in traditional computer systems users appreciate the availability of \ivUndoTechnique{}s, as it helps them recover from mistakes and revert unwanted changes in the \ac{VR} environment. \ivUndoTechnique{}s like the \ivWorldUndo{}, which allow to undo all changes in the \ac{VR} environment independent of the source of the change, increases the connection between users in the same \ac{VR} environments, but also leads to higher reciprocal disturbance of the users.


\section{Related Work}
\label{sec:relatedwork}
To situate our work, we give an overview of different undo mechanics established in traditional computer systems and summarize the state of collaboration in \ac{VR} environments.

\begin{figure*}[ht!]

    \begin{minipage}[t]{.32\linewidth}
		\includegraphics[width=0.96\linewidth]{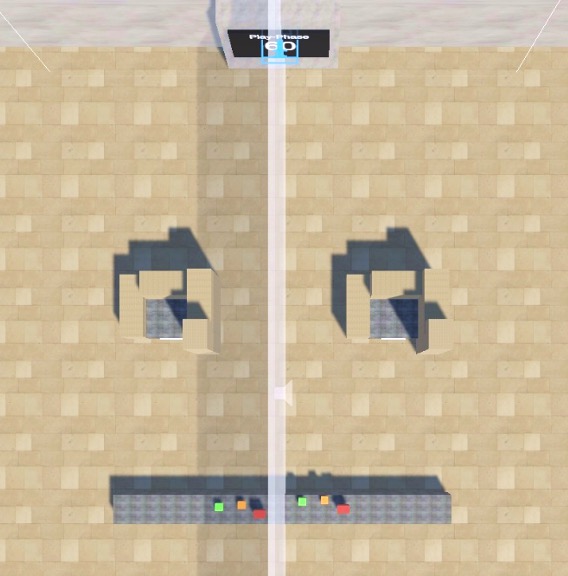}
        \centering
		\subcaption{Top view for \ivDivided{} task{}}
        \label{fig:task_top_div}
	\end{minipage}%
    \begin{minipage}[t]{.32\linewidth}
		\includegraphics[width=0.96\linewidth]{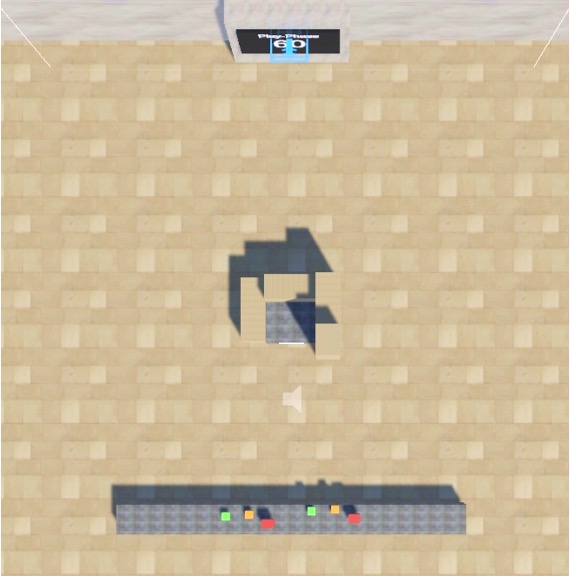}
        \centering
		\subcaption{Top view for \ivCollaborative{} task }
        \label{fig:task_top_col}
	\end{minipage}%
    \begin{minipage}[t]{.36\linewidth}
		\includegraphics[width=0.95\linewidth]{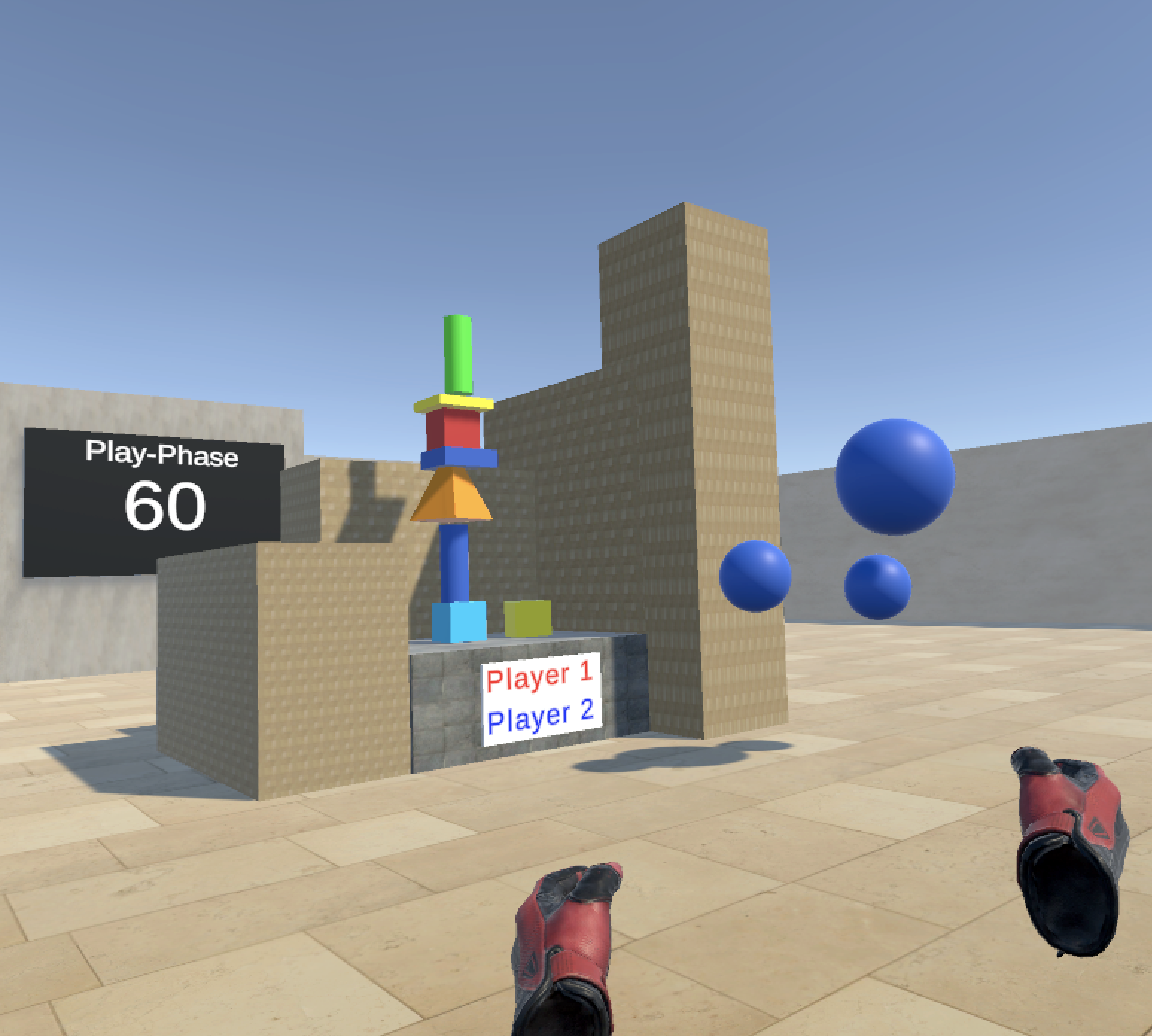}
        \centering
		\subcaption{Participants view for \ivCollaborative{} task }
        \label{fig:task_pov_col}
	\end{minipage}%
 
	\caption{The task environment of our user study. (a) Top view of the task environment for the \ivDivided{} task. A transparent wall separates the game area to prevent interference between two players. Each player has their own stacking area as well as spawn area of the building blocks. (b) Top view of the task environment for the \ivCollaborative{} task. The players share one stacking and block spawning area. (c) Participants perspective during the \ivCollaborative{} task. The center of the image shows the stacking area, on which participants have to place the building blocks. The stairs around the stacking area can be used by participants to reach the top of the tower also for increased heights. The avatar of the other player is represented by one sphere for the head and two smaller spheres for the hands. In the back a "screen" displays the remaining time to construct the tower. Not in the picture is the area to pick up the building blocks. In this case, participants already stacked 7 blocks successfully.}
	
    \Description[Figure showing the task areas for the two different collaboration modes from the top and the participants perspective for the collaborative mode.]{A three part figure showing the task environment. The first two images show a top view of the environment. The main components are the stacking area to place the building blocks, the block spawning area where new blocks appear and a "screen" for the timer. The left subfigure shows the top view of the task environment for the divided task. A transparent wall separates the game area to prevent interference between two players. Each player has their own stacking area as well as spawn area of the building blocks. The middle subfigure shows the top view of the task environment for the collaborative task. Here the players share one stacking and block spawning area. The right subfigure shows the participants perspective of the environment for the collaborative case. The center of the image shows the stacking area, on which participants have to place the building blocks. The stairs around the stacking area can be used by participants to reach the top of the tower also for increased heights. The avatar of the other player is represented by one sphere for the head and two smaller spheres for the hands. In the back a "screen" displays the remaining time to construct the tower. Not in the picture is the area to pick up the building blocks. In this case, participants already stacked 7 blocks of different color and shape successfully}
	\label{fig:TaskEnvironment}
\end{figure*}



\subsection{Undo Actions in Collaborative Systems}



Today, undo mechanics are one of the standard features of many computer systems and are recommended for well-designed \acp{UI} of all kinds \cite{shneiderman_designing_2005}. The feature is present in traditional desktop computers, same as in mobile devices \cite{shima_airflip-undo_2015, loregian_undo_2008}. \citet{teitelman_automated_1972} introduced one of the first instances of an undo mechanic as a feature for the \textit{programmer's assistant} more than 50 years ago. 17 years after this, \citet{yang_undo_1988} states that \nquote{Most sophisticated interface systems should be provided with an undo support} and identifies the undo mechanic as one of three core support features for user interfaces to recover from errors and unwanted situations, besides stops and escapes. 
Further, \citet{myers_taxonomies_1990} work on visual programming paradigms points out the relevance of undo mechanisms in \acp{GUI}. 

For multi-user systems \citet{abowd_giving_1992} emphasize the need for undo support in the context of synchronous group editors, where multiple streams of activity can easily induce errors. In this context \citet{abowd_giving_1992} discuss emerging problems regarding different roles and ownership with respect to objects in the system and identify two models of undo for multi-user systems. A \textit{local undo} only affects changes made by the users themselves, and a \textit{global undo} affects all changes made to the system. Extending on this, \citet{prakash_undoing_1992} propose a \textit{selective undo}, allowing to apply undo only to certain objects in the system. The associated selection can be based on any attribute of the objects, e.g., the identity of the previous user, the manipulation time, or the region of the object. In their subsequent work \citet{prakash_framework_1994} propose the \textit{region undo} besides the per-user history undo and a multiple-operation undo as three practical examples for their \textit{selective undo}. 
Also, more recent work, e.g., by \citet{cass_empirical_2006} advocates for the use of selective undo mechanisms to allow users to also undo actions in a non-linear fashion and maintain dependencies. 
In the context of large interactive surfaces, \citet{seifried_regional_2012} investigate different undo techniques for co-located collaborative workspaces on interactive screens. Here, as well, the authors resort to the three established undo concepts \textit{global}, \textit{personal}, and \textit{selective or regional undo}, each with a different range of effect. These examples underline that undo mechanics are well established and researched in many computer systems, however, their applicability and benefit for \ac{VR} systems remains unexplored.

\subsection{Collaboration in VR}



In recent years, we have seen increasing use of \ac{VR} as a foundation for novel collaborative multi-user experiences in research and industry. This resulted in a large variety of collaborative \ac{VR} applications from entertainment \cite{mcveigh-schultz_shaping_2019, wienrich_social_2018} to professional usage \cite{goebbels_co-presence_2001} and groupware\,\cite{ellis_groupware_1999, gunkel_virtual_2018}. Designers and Engineers use collaborative \ac{VR} applications to iterate over designs in \ac{CAD} application \cite{lehner_distributed_1997, stark_towards_2010, mahdjoub_collaborative_2010} and Collaborative Learning Environments \cite{jackson_collaboration_2000,peng_exploring_2021} offer new ways to engage with learning content. While some applications focus more on transferring the workflow from 2D screens to \ac{VR}, others make use of the new possibilities, e.g., by changing the size of the collaborating users\,\cite{xia_spacetime_2018}. 
Other research explores the use of multi-user \ac{VR} in the context of psychotherapy \cite{matsangidou_now_2022, paping_explorative_2010, triandafilou_development_2018} or for collaborative immersive visual analysis of multidimensional data \cite{butscher_clusters_2018}. Lastly, there are multiple examples for multi-user \ac{VR}-trainings from the industrial \cite{de_paolis_multi-user_2018} and medical domain \cite{schild_applying_2018,lerner_immersive_2020}. 
As a reoccurring pattern, here we can observe two main collaboration types known from traditional computer systems again: divided and collaborative work. These two extremes on an actual continuous spectrum are used, e.g., by \citet{xia_spacetime_2018} and \citet{pinho_cooperative_2008} as a representative distinction for different modes of \ac{VR} collaboration.


Many of the interaction techniques to support collaboration in traditional multi-user collaboration tools, such as synchronized pointing, selecting, and manipulating, made the leap into the \ac{VR} domain already \cite{pinho_cooperative_2008}. While these techniques have been thoroughly studied and are now part of many collaborative \ac{VR} systems in the industry, undo mechanics in these systems are notably absent to a large part. This is in stark contrast to participants in \ac{VR} user studies expressing their desire for undo techniques if not present~\cite{peng_exploring_2021} and researchers encouraging future work to investigate undo mechanics~\cite{feeman_exploration_2018} for \ac{VR} as well.

Recently, research started exploring aspects of undo mechanics in \ac{VR}, in order to revisit questions from \ac{CSCW} and the adaptability to the domain of \ac{VR}. \citet{zhang_vrgit_2023} propose a git-like version control system for collaborative content creation in \ac{VR} to identify similarities and differences to desktop-based systems. \citet{friedman_method_2014} implement a \ac{VR} time travel experience to study users’ responses to the illusion of time travel in the context of a virtual trolley problem \cite{foot_problem_1967}.
And \citet{muller_undoport_2023} investigate the effects of an undo mechanic for point\&teleport-steps on the user behavior in an explorative navigation task. Further, individual single-user \ac{VR} applications like, e.g., \textit{Tilt Brush}\footnote{https://www.tiltbrush.com} provide undo features. 

Beyond the specific focus on \ac{VR} interfaces, previous work also points out the need for undo actions in \ac{AR}~\cite{de_paiva_guimaraes_checklist_2014}. Here, \citet{hutchison_user-defined_2013} discuss potential gestures for \ac{AR} interfaces for 40 common tasks in \ac{AR} including the undo function and \citet{kaufmann_collaborative_2003} maps undo functions to a hand-held tracked panel in an \ac{AR} learning environment.

While these works are exciting and deliver important contributions to the community, they do not offer insights into the transferability of established undo mechanics to \ac{VR} applications. In particular, to the best of our knowledge, there is no prior work on undo mechanics for multi-user \ac{VR} applications. This is supported by \citet{ens_revisiting_2019}, who investigated the status of \ac{MR} groupware and did not find the term \emph{undo} a single time.

With the increasing number of collaborative multi-user applications for \ac{VR} on the one hand and the established effectiveness of undo mechanics as a support feature for traditional (collaborative) computing systems on the other hand, we see the need for a systematic understanding of the transferability of undo actions to the domain of collaborative \ac{VR}.
To assess which characteristics of traditional undo mechanics can or cannot be adopted to these use cases is a complex and nontrivial endeavor as it requires taking into account the intricacies of multi-user VR applications. As one key element unique to \ac{VR}, we see the blending of users' actions to the computer system and changes in the \ac{VR} environment itself, resulting in a strong intertwining of the two. Consequently, undo actions of users result in a change in the users' surroundings. With this work, we aim to contribute to a better systematic understanding of characteristics of undo actions in \ac{VR}.

\begin{figure*}[tb!]
	\begin{minipage}[t]{.5\linewidth}
		\includegraphics[width=0.8\linewidth]{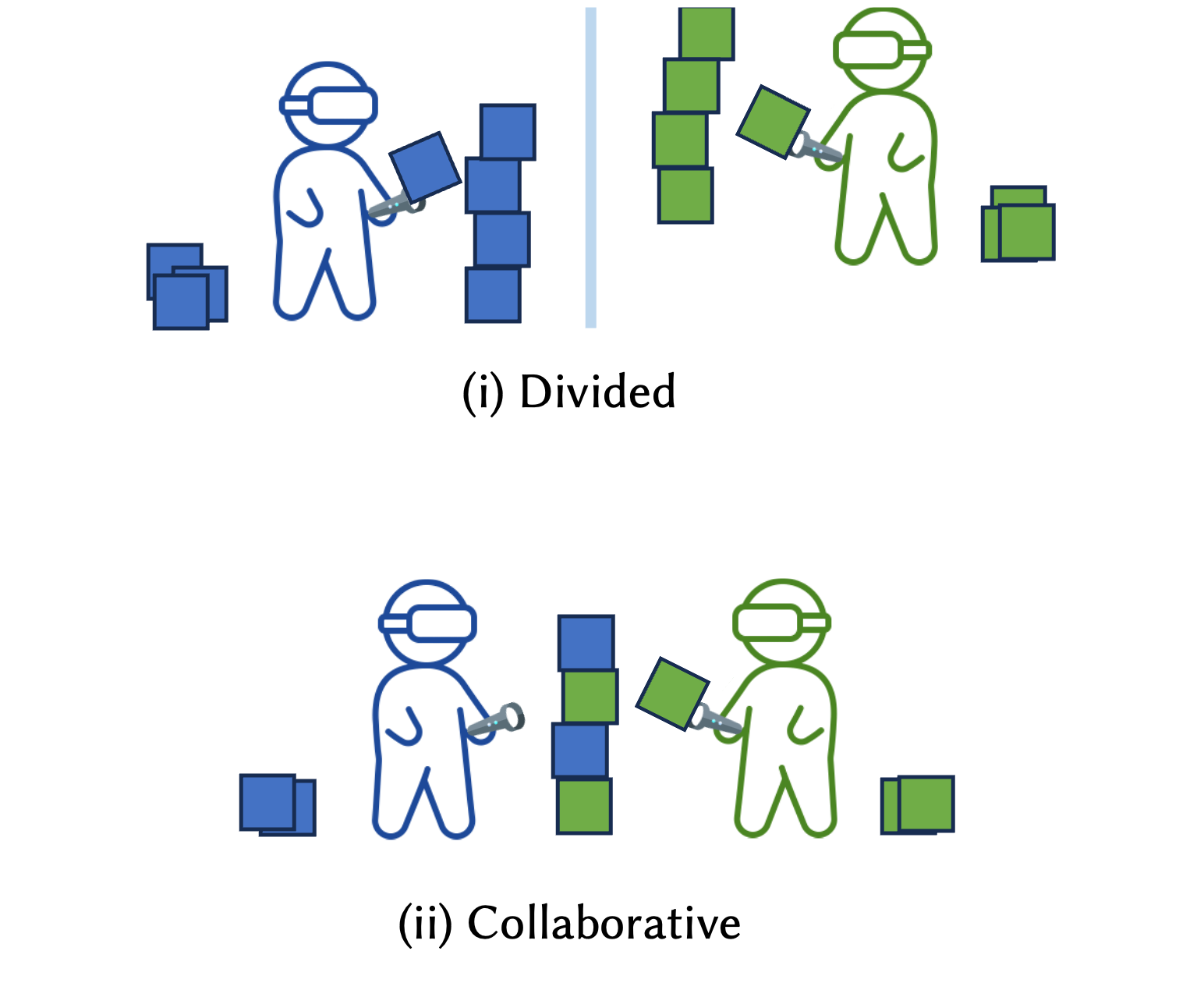}
        \centering
		\subcaption{\ivCollaborationMode{}}\label{fig:IV_ColMode}
	\end{minipage}%
    \begin{minipage}[t]{.5\linewidth}
		\includegraphics[width=0.8\linewidth]{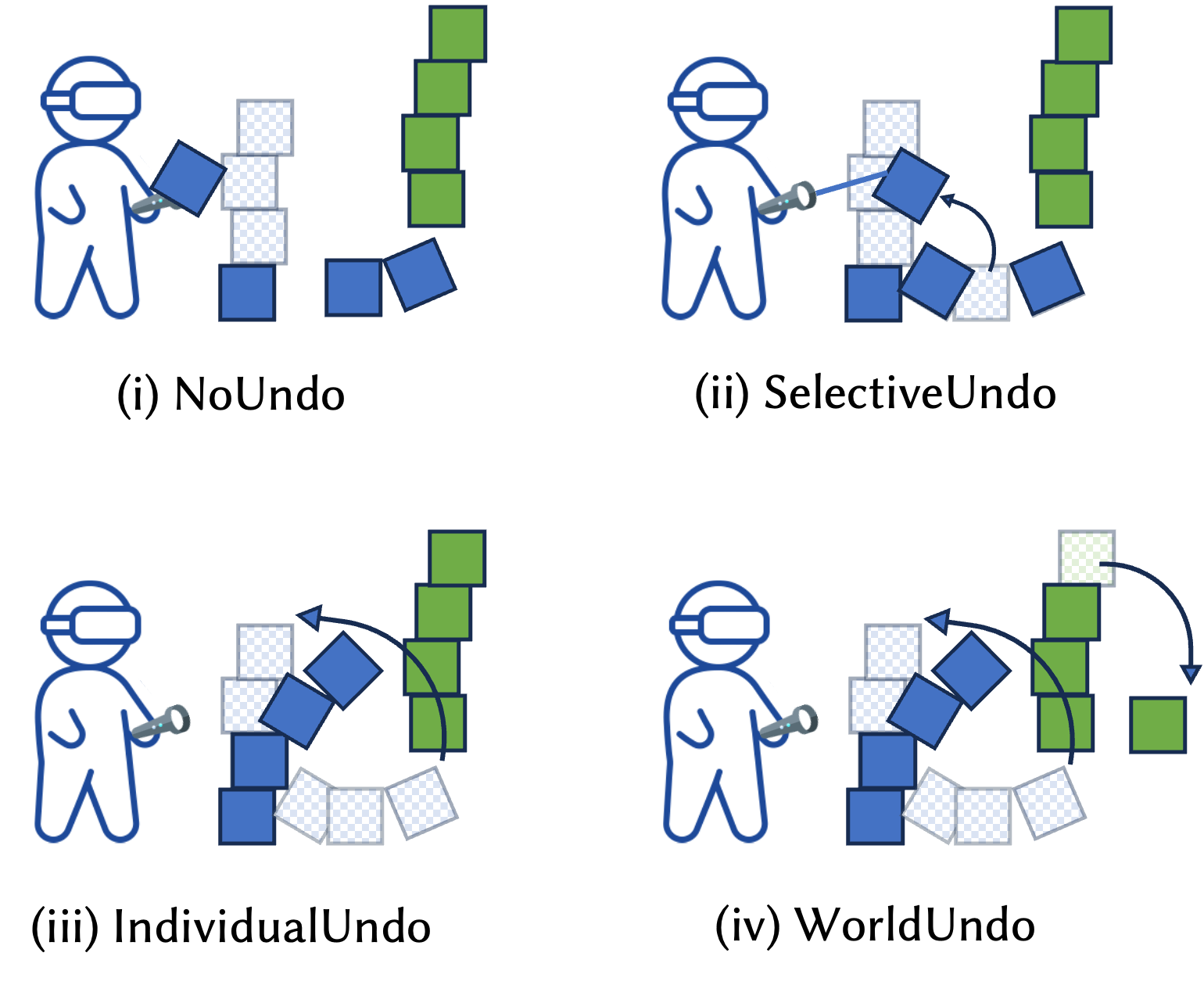}
        \centering
		\subcaption{\ivUndoTechnique{}}\label{fig:IV_UndoTechniques}
	\end{minipage}%

 \caption{(a) The 2 levels of the independent variable \ivCollaborationMode{}s with the (i) \ivDivided{} case on top and the (ii) \ivCollaborative{} case bottom. (b) The 4 levels of the independent variable \ivUndoTechnique{} with the baseline condition (i) \ivNoUndo{} as well as the 3 different \ivUndoTechnique{}s (ii) \ivSelectiveUndo{}, (iii) \ivIndividualUndo{} and (iv)\,\ivWorldUndo{}.}

 \Description[The independent variables varied in the experiment.]{A figure consisting of 2 subfigures, explaining the two independent variables. The left subfigure consists of two images, top and bottom. Each depicts two people wearing a VR headset stacking cubes in two different collaboration modes. On the top, the two people are separated by a wall and building two separate towers. On the button the two people are collaboratively constructing a joint tower. The right subfigure consists of four images and depicts a single blue person utilizing the four different undo techniques. The top left shows a collapsed blue tower and standing green tower and has the caption "(i) NoUndo". The top right shows a single block of the collapsed tower being moved back to its position before the collapse and a standing green tower. A ray point towards this block, the caption is "(ii) SelectiveUndo". The bottom left image shows all blocks of the collapsed blue tower transitioning back to their previous position and a standing green tower. The caption is "(iii) IndividualUndo". The bottom right tower is identical to the previous, but not from the standing green tower one block is removed back to the ground. The caption is "(iv) WorldUndo"}
	\label{fig:IndependendVariables}
\end{figure*}

\section{Methodology}
\label{sec:methodology}


The analysis of previous work (see \autoref{sec:relatedwork}) revealed that undo techniques are an established and well-researched means in traditional computer systems to support individual and collaborative activities. On the other hand, we found a surprising lack of systematic insights into the use and best practices of undo techniques for VR, especially for collaborative applications. Based on the findings from related work and current practices in \ac{VR}, we formulate the following research hypotheses:

\begin{description}[labelindent=1pt]
    \item[H1:] The availability of undo in \ac{VR} increases users' performance.
    \item[H2:] The availability of undo in \ac{VR} increases the user experience.
    \item[H3:] Undo techniques that affect all users increase the social connectedness of users in \ac{VR}.
    \item[H4:] Undo techniques that affect all users increase the interference of users during non-collaborative work in \ac{VR}.
\end{description}


Many collaborative interaction techniques from traditional computer systems, like synchronized pointing, selecting, and manipulating, are adopted to collaborative \ac{VR} already \cite{pinho_cooperative_2008}. With H1 and H2, we investigate if the implementation of undo mechanics as a support feature for collaborative \ac{VR} applications can assist users similar to traditional (collaborative) computer systems \cite{shneiderman_designing_2005}. With H3 and H4, we study the effects different established forms of undo\,\cite{abowd_giving_1992, prakash_undoing_1992} have on the relation between two users.

To test our hypotheses, we explore the influence of 1) the mode of collaboration between users to account for different task scenarios common to co-located collaborative work \cite{xia_spacetime_2018, pinho_cooperative_2008, seifried_regional_2012} and 2) the range of effect of the undo technique as established in \ac{CSCW} \cite{abowd_giving_1992,prakash_undoing_1992}.







\subsection{Design and Task}
For our user study, we designed a construction task for two participants in \ac{VR}.
As the canvas for the undo techniques, we selected three-dimensional primitive shapes as suitable artifacts representative for the three-dimensionality of the \ac{VR} domain. These afford all interactions inherent to \ac{VR} and shape the environment itself. 
We instructed the participants to build a tower with blocks of different basic 3D geometries while being in the same virtual space as another participant, as shown in \autoref{fig:TaskEnvironment}. To account for different scenarios, participants worked on the task either together or alone, depending on the condition. When working alone, we adjusted the scene as shown in \autoref{fig:task_top_div}, to provide a stacking area for each participant and placed a transparent wall separating the game space of the virtual world to avoid unintended interference between the participants.
In both cases, the goal is to build a tower as high as possible within a task time of 4\,minutes. We chose this time frame as it allowed participants enough time to reach a critical stability of the tower while limiting the total study time to a maximum of 90\,minutes. During the task participants moved the blocks from a spawning location to a target location, where they stacked them. Once they placed a block on the tower, a new block appeared in the spawning location, ensuring sufficient blocks throughout the condition. With increasing height of the tower, the towers' stability decreases naturally, resulting in an eventual collapse. We informed participants that the height of the tower at the end of the task time is relevant to ensure participants approach the task carefully.
We chose a minimalistic avatar design to visualize the users to reduce distractions, and encourage participants to focus on the task.
We opted for a controller-based input device for the input modality as the current de facto standard for \ac{VR} interactions.
We chose this task for the following reasons. First, it uses the standard \ac{VR} manipulation techniques (translate, rotate, grab, release) and requires locomotion. Second, the task can be solved individually as well as collaboratively. Third, it changes the environment of the participant. And fourth, it allows us to measure the progress and success of participants.
The study design was approved by our institution's ethics committee. 

\subsection{Independent Variables}
As described at the beginning of this section, we investigate the effects of the \ivCollaborationMode{} and available \ivUndoTechnique{} to discover the influence on the dependent variables. We vary these two independent variables with the following levels:

\begin{description}

    \item[\ivCollaborationMode{}:] To attribute for different modes of collaboration, we chose two levels as shown in \autoref{fig:IV_ColMode}:
    \begin{description}[labelindent=1pt]
    \item[\textit{\ivDivided{}}] Players build their individual tower in their separate game space.
    \item[\textit{\ivCollaborative{}}] Players build a tower together in a joint game space.
    \end{description}
\vspace{8pt}
    
    \item[\ivUndoTechnique{}:] Based on the concept described before, we investigate three different levels of undo functions and a baseline. The \ivUndoTechnique{}s are shown in \autoref{fig:IV_UndoTechniques} (i)-(iv):
    \begin{description}[labelindent=1pt]
    \item[\textit{\ivNoUndo{}}] This level represents the current de facto standard VR interactions without any undo function.
    \item[\textit{\ivSelectiveUndo{}}] Players can undo object manipulations of the object selected by the ray interactor.
    \item[\textit{\ivIndividualUndo{}}] Players can undo their own object manipulations in a linear way.
    \item[\textit{\ivWorldUndo{}}] Players can undo all object manipulations in the scene in a linear way, also affecting the actions of the other player.
    \end{description}

\end{description}

\noindent
Based on the presented levels of the two independent variables, participants experience 8 different conditions throughout the study. To prevent learning effects and reduce order-effects and carry-over-effects between the conditions, we counterbalance the conditions using the \emph{balanced latin square design}.

We chose not to include the respective other user's avatar in any range of effect for the following reasons. First, we want not to objectify users. Second, we want to maintain individuals' freedom of movement as users do not like to be moved by others' \cite{rasch_going_2023}. Third, as mentioned before, the aspect of locomotion is covered already \cite{muller_undoport_2023}, although only in a single-user context. This results in a 2\,x\,4 space shown in \autoref{fig:IndependendVariables}, which we investigate in the context of our controlled experiment, described in more detail in the following.


\subsection{Dependent Variables}
To assess the influence of the \ivCollaborationMode{} and \ivUndoTechnique{}, we survey the influence on the following measures: 
\begin{description}
    \item[\dvIOS{} Score] The \ac{IOS} is a single-item questionnaire proposed by \citet{aron_inclusion_1992} and measures the perceived psychological closeness to other persons or groups. With this measure, we aim to study H3.
    \item[\dvGEQ{} Score] The Behavioural Involvement Component of the Social Presence Module in the \ac{GEQ} by \citet{ijsselsteijn_game_2013} measures how much attention participants pay to their counterpart. The mentioned component uses six of the 17 items of the Social Presence Module of the \ac{GEQ} and uses a five-point Likert scale for answers. We include this measure to study H3 and H4.
    \item[\dvTLX{}] The \ac{RTLX} \cite{hart_nasa-task_2006} measures how demanding each condition is. The six items of the \ac{RTLX} use a 0-100 scale for answers to study H1 and H2. 
    \item[Custom Questionnaire] The questionnaire on an established five-point Likert scale holds eight items regarding the feeling of \dvDisturbing{}, \dvDisturbed{}, \dvControl{}, \dvFrustration, \dvSuccess{}, \dvRecover{}, \dvEnjoyment{} and \dvDesiredFutureUsage{}. The results section presents the complete statement for each item. We chose this custom questionnaire in addition to the established questionnaires to allow for specific insights for H1, H2, and H4 for the study on hand.
\end{description}

\noindent
Besides the survey data, we also log the users' interactions throughout each condition to assess the influence of the \ivCollaborationMode{} and \ivUndoTechnique{}. During the experiment, we logged the following data:
\begin{description}
    \item[\dvNumberGrabs{}] The total number of grab interactions a participant performed during a condition as an efficiency and engagement measure. With this measure, we want to understand if the availability of an undo feature affects the frequency of the conventional grab interaction and study H1.
    \item[\dvNumberUndos{}] The total number of undo operations a participant performed during a condition as a measure of participants' acceptance and usage. With this measure, we want to understand if participants use an undo feature if available and study H1, H2, and H4.
    \item[\dvTowerHeight{}] The final height of the constructed tower in meters at the end of a condition as a measure of the performance. With this measure, we aim to detect changes in participants' performance on the task for different undo features and study H1 and H4.
\end{description}    

We excluded time as an efficiency measure, as we needed a fixed time per condition to study the effects on participants' willingness for last-minute changes and create time pressure towards the end of each condition.

\begin{figure*}[th!]
    \vspace{2em}
	\includegraphics[width=\linewidth]{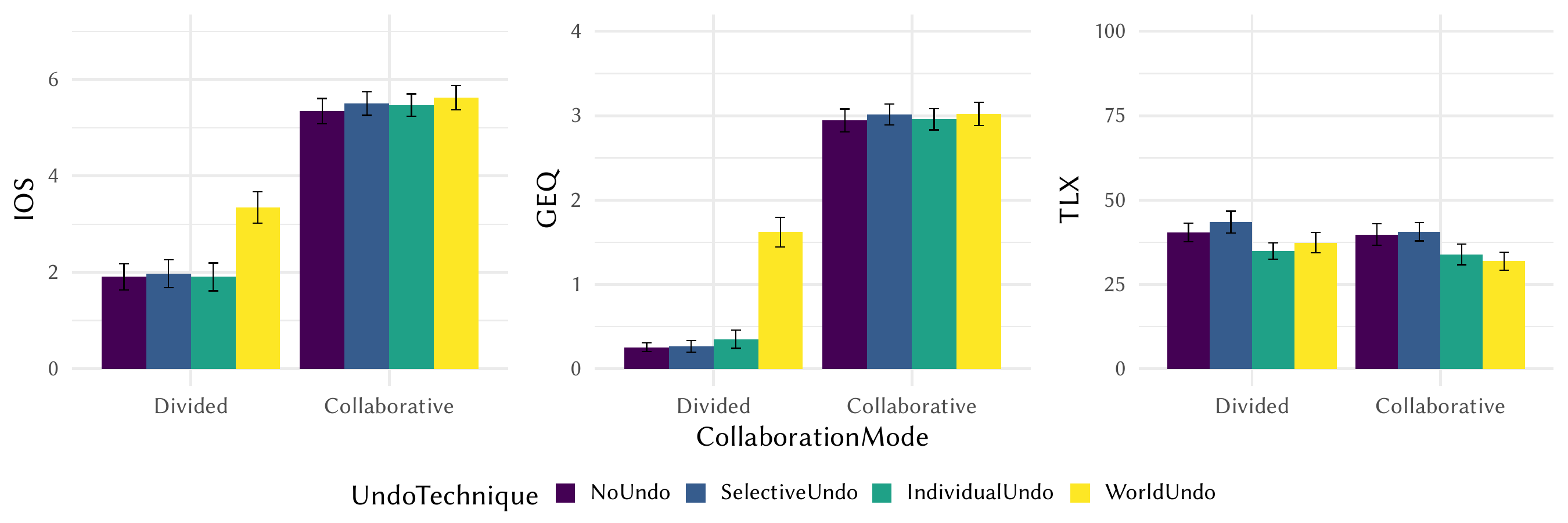}

    \vspace{-1em}
	\begin{minipage}[t]{.3\linewidth}
		\centering
		\subcaption{\dvIOS{} score}\label{fig:results:IOS}
	\end{minipage}%
    \begin{minipage}[t]{.3\linewidth}
		\centering
		\subcaption{\dvGEQ{} score}\label{fig:results:GEQ}
	\end{minipage}%
     \begin{minipage}[t]{.3\linewidth}
		\centering
		\subcaption{\dvTLX{} score}\label{fig:results:TLX}
	\end{minipage}%
 
	\caption{The mean results for (a) \dvIOS{}, (b) \dvGEQ{}, and (c) \dvTLX{} as a bar chart plot. The error bars indicate the standard error.}
	\Description[The graphs for the IOS, GEQ, and TLX measure]{A figure consisting of three bar charts. Each plot is separated into the Divided and Collaborative section. Each of these section depicts the results for the four undo conditions, NoUndo SelectiveUndo, IndividualUndo, WorldUndo. The first plot shows the IOS rating, the second the GEQ score, and the third the RTLX score.}
	\label{fig:IOS_GEQ_TLX}
\end{figure*}

\subsection{Apparatus}

We implemented the virtual environment and interaction techniques as detailed above as a multi-player \ac{VR} application and deployed it on two connected computers with two \ac{HMD}s using a shared tracking space. %
We developed the study design and its functionalities using the Unity Version 2021.3.14f. We used two HTC Vive Pro as VR devices, each consisting of a \ac{HMD}, two HTC motion controllers, and two base stations for tracking. We executed the application on two VR computers with the following specifications: Intel Core i7-10700, 16\,GB RAM, NVIDIA GeForce RTX 2060 SUPER. %
For connecting the individual users' applications, we used the multi-player networking framework Photon Pun\footnote{https://www.photonengine.com/en-US/Photon} for Unity, which supports room creation, matchmaking, and event-based communication for real-time scenarios. To connect to the internet and minimize networking complications, both computers were connected to the same network via LAN cable.

To implement the \ivUndoTechnique{}s, we store the position and rotation of each object together with a time-stamp and the manipulating agent in a data frame throughout the runtime of the \ac{VR} application. When a participant uses the undo feature, we access the previous states of objects and assign the position accordingly. For the \ivSelectiveUndo{}, we use a ray interactor for selection. Depending on the current condition, we access all objects, the selected object, or all objects manipulated by one player from the data frame.

We used a total area of 3.2 x 5.5 meters as physical space and set up two individual VR spaces with active Steam Guard to prevent collision with walls or the other participant.

\subsection{Procedure}

After welcoming the participants, we provide an overview of the study's objectives and their tasks. Participants are then asked to sign a consent form to authorize the collection and processing of their data. They then complete a pre-survey covering demographics, experience with virtual reality, and familiarity with the other participants.
Next, we demonstrate fundamental VR controls for teleportation and object interaction, allowing participants a few minutes to practice before the study begins. We then explain their primary task in VR, namely constructing a tower using building blocks within a four-minute time frame, aiming to maximize its height. We emphasize that building blocks should remain stationary when time expires to contribute to their score. We then let participants enter a tutorial scene to allow them to familiarize themselves with basic controls, ensuring they understand the task and are able to perform it. Additionally, before each test condition, participants have one minute to familiarize themselves with the provided undo function of the condition. Throughout the conditions, participants can talk with each other. After each of the eight conditions, they are asked to complete a questionnaire on a PC. After finishing all conditions, participants take a short break before filling out a post-study survey to evaluate and provide feedback on the undo functions relative to each other. After this, we collect further qualitative feedback from the participants. Finally, we provide compensation for their participation. Throughout the study of 90\,minutes participants experience 8 different conditions for 4\, minutes each in a counterbalanced order.

\subsection{Participants}
We recruited 32 participants (18 female, 14 male) with a mean age of 26.56, ranging from 19 to 58. Of the participants, 18 knew the other participant, while 14 did not know the other participant. 10 participants stated to be a \textit{experienced VR user}, 10 stated to be a  \textit{sporadic VR user} and 12 had \textit{no VR Experience} at all.


\subsection{Analysis}
For analyzing the non-parametric data, we applied the \ac{ART} proposed by \citet{wobbrock_aligned_2011}. For significant results, we follow the ART-C procedure suggested by \citet{elkin_aligned_2021}. %
We report the generalized ETA squared \ges{} as a measure of the effect and classify it in alignment with \citet{bakeman_recommended_2005}. Here, we use suggestions by \citet{cohen_statistical_2013} for small ($>.0099$), medium ($>.0588$), or large ($>.1379$) effect size.

For analyzing the parametric data, we tested the data with Shapiro-Wilk's and Mauchly's tests for normality of the residuals and sphericity assumptions. When the assumption of sphericity was violated, we used the Greenhouse-Geisser method to correct the tests.
We used two-way repeated-measures ANOVAs to identify significant effects and applied Bonferroni-corrected t-tests for post-hoc analysis. If normality was violated, we performed a non-parametric analysis as described before.

To analyze the count data from the log files, we fitted Poisson regression models and applied Type III Wald chi-square tests for significance testing. Here, we used the Tukey method for p-value adjustments.

\section{Results}
\label{sec:results}


This section presents the results of our controlled experiment.

\begin{figure*}[t!]
	
	\centering
	\includegraphics[width=\textwidth]{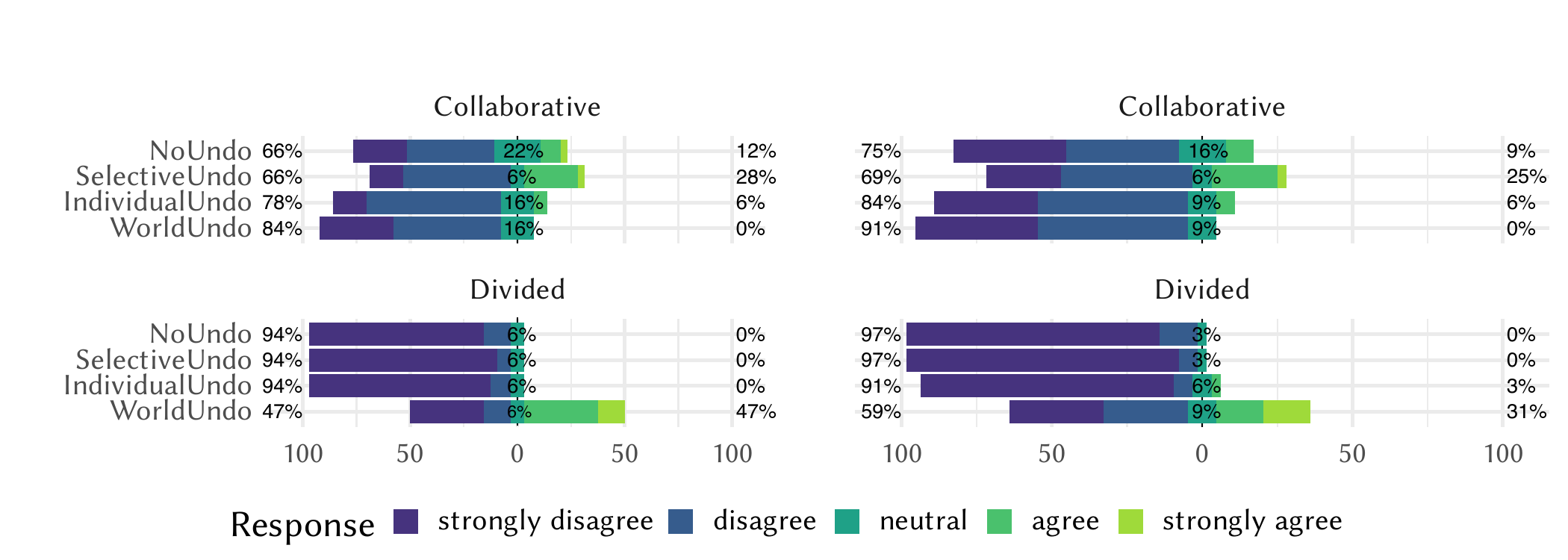}\hfill
	\vspace{-1em}
	\begin{minipage}[t]{.5\linewidth}
		\centering
		\subcaption{\dvDisturbing}\label{fig:results:disturbing}
	\end{minipage}%
	\begin{minipage}[t]{.5\linewidth}
		\centering
		\subcaption{\dvDisturbed}\label{fig:results:disturbed}
	\end{minipage}%
	
	\caption{Participants' responses regarding (a) \dvDisturbing{} and (b) \dvDisturbed{} on a 5-point Likert scale. The percentage number indicates the proportion of the answers for negative, neutral, and positive responses.}
    \Description[Four Likert plots showing the participants responses to "disturbing others" and "felt disturbed"]{Four Likert plots showing the participants responses to "disturbing others" and "felt disturbed". The two left graphs show the responses for DisturbingOthers Collaborative (top) and Divided (bottom). The two right graphs show the responses for FeltDisturbed Collaborative (top) and Divided (bottom) Each plot consists of four subplots with the responses per technique on a five point Likert scale.}
	\label{fig:results:12}
\end{figure*}

\subsection{IOS Score}
For the single-item \ac{IOS} questionnaire, we took the participants' ratings and compared them directly as proposed by \citet{aron_inclusion_1992}. %
We analyzed the \ac{IOS} score as a measure of social connectedness between participants.
Here, we found mean values ranging from $M= 5.62, SD = 1.45$ (\ivCollaborative{}, \ivWorldUndo{}) to M= 1.91, SD = 1.63 (\ivDivided{}, \ivIndividualUndo{}) shown in \autoref{fig:IOS_GEQ_TLX}. The \ac{ART} ANOVA showed a significant (\ano{1}{31}{146.70}{<.001}) main effect for the \ivCollaborationMode{} on the \ac{IOS} scale with a \efETAsquared{0.82} effect size. Post-hoc tests revealed significantly  ($p<.001$) higher ratings for \ivCollaborative{} ($M=5.48, SD=1.40$) compared to \ivDivided{} ($M=2.28, SD=1.76$).

We also found a significant (\ano{3}{93}{12.52}{<.001}) main effect for the \ivUndoTechnique{} on the \ac{IOS} scale with a \efETAsquared{0.28} effect size. Post-hoc tests revealed significantly higher ratings for \ivWorldUndo{} ($M=4.48 , SD=2.01$) compared to all other levels (\ivNoUndo{}: $M = 3.62, SD = 2.29, p<.001$, \ivSelectiveUndo{}: $ M = 3.73, SD = 2.32, p<.001$, \ivIndividualUndo{}: $ M = 3.69, SD = 2.32, p<.01$).

Further, we found interaction effects (\ano{3}{93}{5.56}{<.01}) with a \efETAsquared{0.15} effect size. While we could not find differences between the \ivUndoTechnique{}s in the \ivCollaborative{} case (all $p>.05$), \ivWorldUndo{} received significantly higher ($p<.001$) ratings compared to all other levels  in the \ivDivided{} case.

These results support H3, showing that \ivUndoTechnique{}s with a global range of effect like \ivWorldUndo{} can increase the social connectedness of the users.


\subsection{GEQ Score} 

For the \ac{GEQ}, we evaluated the Behavioral Involvement Component as an average of its items, according to the scoring guideline \cite{ijsselsteijn_game_2013}.
Evaluating the Behavioral Involvement Component of the \ac{GEQ} Social Presence Module, the analysis yielded values ranging from $M = 3.02, SD = 0.78$  (\ivCollaborative{}, \ivWorldUndo{}) to $M= 0.25, SD = 0.3$ (\ivDivided{}, \ivNoUndo{}).

Using the \ac{ART} ANOVA, we found a significant (\ano{1}{31}{358.836}{<.001}) main effect for the \ivCollaborationMode{} on the \ac{GEQ} rating with a \efETAsquared{0.92} effect size. Post-hoc tests revealed significantly ($p<.001$) higher ratings for \ivCollaborative{} ($M = 2.98, SD = 0.74$) compared to \ivDivided{} ($M = 0.62, SD = 0.85$).
We also found a significant (\ano{3}{93}{18.09}{<.001}) main effect for the \ivUndoTechnique{} on the \ac{GEQ} rating with a \efETAsquared{0.36} effect size. Here, post-hoc tests revealed significantly ($p<.001$) higher ratings for \ivWorldUndo{} ($M = 2.32, SD = 1.14$) compared to all other levels. (\ivNoUndo{}: $M = 1.60, SD = 1.47$, \ivSelectiveUndo{}: $M = 1.64, SD = 1.50$, \ivIndividualUndo{}: $M = 1.65, SD = 1.47$).

Again, we found a significant (\ano{3}{93}{17.83}{<.001}) interaction effect with a \efETAsquared{0.36} effect size. While we could not find differences between the \ivUndoTechnique{}s in the \ivCollaborative{} case ($p>.05$), \ivWorldUndo{} received significantly ($p<.001$) higher ratings compared to all other levels in the \ivDivided{} case.

These results support H3 and H4, again showing that \ivUndoTechnique{}s with a global range of effect like \ivWorldUndo{} can increase users' social connectedness and mutual interference.


\subsection{Raw TLX Score}
We calculated the \ac{RTLX} score as proposed by \citet{hart_nasa-task_2006}. %
Evaluating the \ac{RTLX}, we found values ranging from $M=43.5, SD = 18.2$ (\ivDivided{}, \ivSelectiveUndo{}) to $M= 31.9, SD = 15.0$ (\ivCollaborative{}, \ivWorldUndo{}). Our \ac{ART} ANOVA showed a significant (\ano{3}{93}{9.34}{<.001}) main effect for the \ivUndoTechnique{} on the \ac{RTLX} score with a \efETAsquared{0.23} effect size. Here, post-hoc tests revealed significantly higher ratings for \ivNoUndo{} ($M = 40.1, SD = 16.8$) compared to \ivIndividualUndo{} ($M = 34.4, SD = 15.5$)  and \ivWorldUndo{} ($p<.01$) ($M = 34.6, SD = 16.1$) as well as significantly higher ratings for \ivSelectiveUndo{} ($M = 42.0, SD = 16.7$) compared to \ivIndividualUndo{} ($p<.001$) and \ivWorldUndo{} ($p<.01$). We could not find a significant main effect (\ano{1}{31}{3.91}{>.05}) for the \ivCollaborationMode{} nor interaction effects (\ano{3}{93}{.41}{>.05}) between the two independent variables.

These results only partially support H1 and H2 since not all \ivUndoTechnique{}s positively affected the performance and user experience. We discuss this in more detail in \autoref{sec:discussion}.

\begin{figure*}[t!]
	
	\centering
	\includegraphics[width=\textwidth]{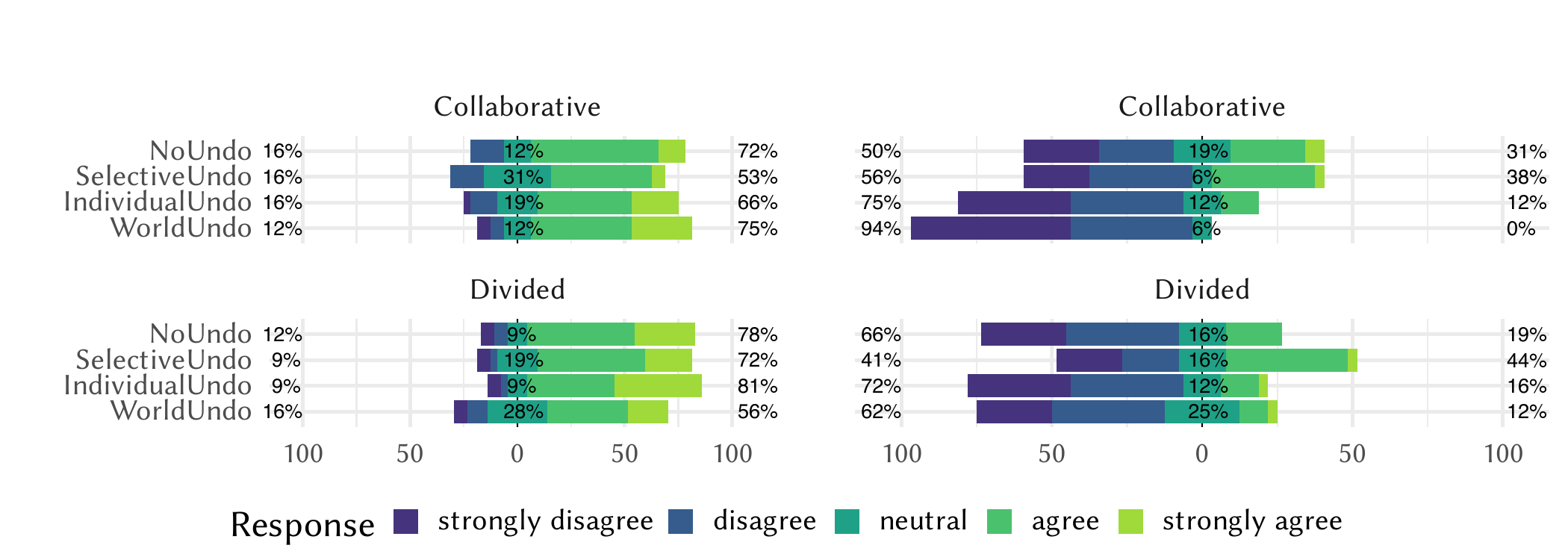}\hfill
	\vspace{-1em}
	\begin{minipage}[t]{.5\linewidth}
		\centering
		\subcaption{\dvControl}\label{fig:results:control}
	\end{minipage}%
	\begin{minipage}[t]{.5\linewidth}
		\centering
		\subcaption{\dvFrustration}\label{fig:results:frustration}
	\end{minipage}%

	\caption{Participants' responses regarding (a) \dvControl{} and (b) \dvFrustration{} on a 5-point Likert scale.}
    \Description[Four Likert plots showing the participants responses to "Control" and "Frustration"]{Four Likert plots showing the participants responses to "Control" and "Frustration". The two left graphs show the responses for Control Collaborative (top) and Divided (bottom). The two right graphs show the responses for Frustration Collaborative (top) and Divided (bottom) Each plot consists of four subplots with the responses per technique on a five point Likert scale.}
	\label{fig:results:34}
\end{figure*}

\subsection{Custom Questionnaire} 
In the following, we present the results of the \ac{ART} ANOVA of our custom questionnaire.

\subsubsection{\dvDisturbing{} "I felt that I disturbed the other player"}
We found a significant (\ano{1}{31}{33.86}{<.001}) main effect for the \ivCollaborationMode{} on the \dvDisturbing{} rating with a \efETAsquared{0.52} effect size. Post-hoc tests revealed significantly higher ratings for \ivCollaborative{} compared to \ivDivided{} ($p<.001$).
We also found a significant (\ano{3}{93}{5.02}{<.01}) main effect for the \ivUndoTechnique{} on the \dvDisturbing{} rating with a \efETAsquared{0.13} effect size. Post-hoc tests revealed significantly higher ratings for \ivWorldUndo{} compared to \ivNoUndo{} ($p<.05$) and \ivIndividualUndo{} ($p<.01$).

Further, we found an interaction effect (\ano{3}{93}{24.07}{<.001}) with a \efETAsquared{0.43} effect size.  While we could not find differences between the \ivUndoTechnique{}s in the \ivCollaborative{} case ($p>.05$), \ivWorldUndo{} received significantly ($p<.001$) higher ratings compared to all other levels in the \ivDivided{} case.

These results support H4, showing \ivUndoTechnique{}s with a global range of effect like \ivWorldUndo{} can increase mutual interference of the users.

\subsubsection{\dvDisturbed{} "I felt that the other player disturbed me"}
We found a significant (\ano{1}{31}{12.86}{<.01}) main effect for the \ivCollaborationMode{} on the \dvDisturbed{} rating with a \efETAsquared{0.29} effect size. Post-hoc tests revealed significantly higher ratings for \ivCollaborative{} compared to \ivDivided{} ($p<.001$).
We also found a significant (\ano{3}{93}{7.42}{<.001}) main effect for the \ivUndoTechnique{} on the \dvDisturbed{} rating with a \efETAsquared{0.19} effect size. Post-hoc tests revealed significantly higher ratings for \ivWorldUndo{} compared to \ivNoUndo{} ($p<.01$) and \ivIndividualUndo{} ($p<.001$).

Again, we found interaction effects (\ano{3}{93}{12.46}{<.001}) with a \efETAsquared{0.28} effect size.  While we could not find differences between the \ivUndoTechnique{}s in the \ivCollaborative{} case ($p>.05$), \ivWorldUndo{} received significantly ($p<.001$) higher ratings compared to all other levels  in the \ivDivided{} case.
 
These results support H4, showing \ivUndoTechnique{}s with a global range of effect like \ivWorldUndo{} can increase mutual interference of the users.

\subsubsection{\dvControl{} "I felt in control over my surroundings"}
We found no significant ($p>.05$) main effect for \ivCollaborationMode{} and \ivUndoTechnique{} on the \dvControl{} rating. However, we found significant (\ano{3}{93}{4.97}{<.01}) interaction effects with a \efETAsquared{0.13} effect size. However, post-hoc tests did not confirm ($p > .05$) significant differences between the groups, and therefore neither contradict nor support H1, H2, and H4.

\subsubsection{\dvFrustration{} "I felt frustrated"}
We found a significant (\ano{3}{93}{7.09}{<.001}) main effect for the \ivUndoTechnique{} on the \dvFrustration{} rating with a \efETAsquared{0.18} effect size. Post-hoc tests revealed significantly higher ratings for \ivSelectiveUndo{} compared to \ivIndividualUndo{} ($p<.01$) and \ivWorldUndo{} ($p<.001$). We could not find a significant main effect (\ano{1}{31}{.21}{>.05}) for the \ivCollaborationMode{} nor interaction effects (\ano{3}{93}{1.55}{>.05}) between the two independent variables.

These results neither contradict nor support H1, H2, and H4.

 \begin{figure*}[t!]
	
	\centering
	\includegraphics[width=\textwidth]{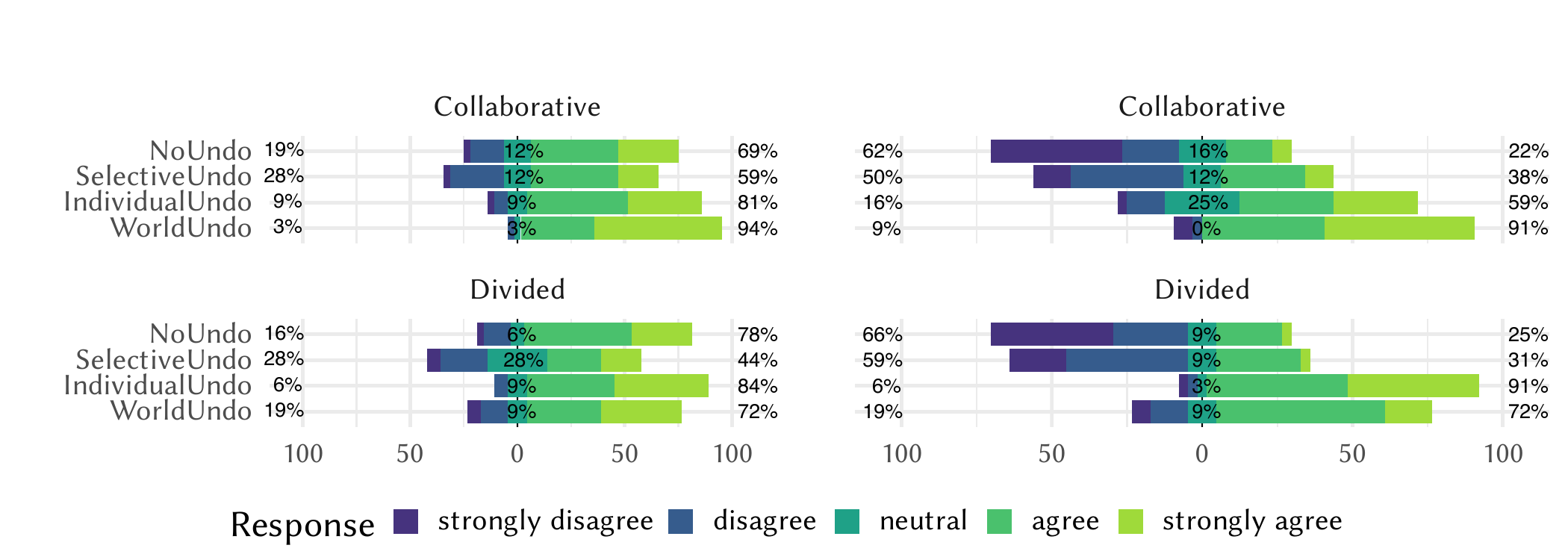}\hfill
	\vspace{-1em}
	\begin{minipage}[t]{.5\linewidth}
		\centering
		\subcaption{\dvSuccess}\label{fig:results:success}
	\end{minipage}%
    \begin{minipage}[t]{.5\linewidth}
		\centering
		\subcaption{\dvRecover}\label{fig:results:recover}
	\end{minipage}%

	\caption{Participants' responses regarding (a) \dvSuccess{} and (b) \dvRecover{} on a 5-point Likert scale.}
    \Description[Four Likert plots showing the participants responses to "Success" and "Recover"]{Four Likert plots showing the participants responses to "Success" and "Recover". The two left graphs show the responses for Success Collaborative (top) and Divided (bottom). The two right graphs show the responses for Recover Collaborative (top) and Divided (bottom) Each plot consists of four subplots with the responses per technique on a five point Likert scale.}
    \label{fig:results:67}
\end{figure*}

\subsubsection{\dvSuccess{} "I felt like I solved the task successfully"}
We found a significant (\ano{3}{93}{7.32}{<.001}) main effect for the \ivUndoTechnique{} on the \dvSuccess{} rating with a \efETAsquared{0.19} effect size. Post-hoc tests revealed significantly higher ratings for \ivIndividualUndo{} and \ivWorldUndo{} compared to \ivSelectiveUndo{} ($p<.001$). Again, we could not find a significant main effect (\ano{1}{31}{.00}{>.05}) for \ivCollaborationMode{} nor interaction effects (\ano{3}{93}{1.67}{>.05}) between the two independent variables.

These results only partially support H2, as not all \ivUndoTechnique{}s show a positive effect on the user experience. We discuss this in more detail in \autoref{sec:discussion}.

\subsubsection{\dvRecover{} "I felt that I could easily recover from my mistakes"}
We found a significant (\ano{3}{93}{32.12}{<.001}) main effect for the \ivUndoTechnique{} on the \dvRecover{} rating with a \efETAsquared{0.50} effect size. Post-hoc tests revealed significantly higher ratings for \ivIndividualUndo{} and \ivWorldUndo{} compared to \ivSelectiveUndo{} ($p<.001$) as well as for  \ivIndividualUndo{} and \ivWorldUndo{} compared to \ivNoUndo{} ($p<.001$). We could not find a significant main effect (\ano{1}{31}{1.63}{>.05}) for \ivCollaborationMode{}, but found a significant (\ano{3}{93}{5.46}{<.01}) interaction effect with a \efETAsquared{0.14} effect size. While in the \ivCollaborative{} case, the \ivWorldUndo{} received higher ratings compared to \ivIndividualUndo{}, in the \ivDivided{} case, this inverts, resulting in higher ratings for \ivIndividualUndo{} compared to the \ivWorldUndo{}. However, these differences are not significant ($p>.05$).

These results partially support H1 and H2, as again, not all \ivUndoTechnique{}s show a positive effect on the performance and user experience. We discuss these results in more detail in \autoref{sec:discussion}.

\subsubsection{\dvEnjoyment{} "I enjoyed using the undo feature"}
We found a significant (\ano{3}{93}{48.05}{<.001}) main effect for the \ivUndoTechnique{} on the \dvEnjoyment{} rating with a \efETAsquared{0.60} effect size. Post-hoc tests revealed significantly higher ratings for \ivIndividualUndo{} and \ivWorldUndo{} compared to \ivNoUndo{} and \ivSelectiveUndo{} ($p<.001$). We could not find a significant main effect (\ano{1}{31}{1.81}{>.05}) for \ivCollaborationMode{} but found significant (\ano{3}{93}{8.67}{<.001}) interaction effects with a \efETAsquared{0.05} effect size. While in the \ivCollaborative{} case \ivWorldUndo{} received non-significant higher ratings compared to \ivIndividualUndo{}, again in the \ivDivided{} case, this inverts, resulting in significantly ($p<.05$) higher ratings for \ivIndividualUndo{} compared to \ivWorldUndo{}. 

These results partially support H2, as they are inconsistent across the \ivUndoTechnique{}s. We discuss this in more detail in \autoref{sec:discussion}.

\subsubsection{\dvDesiredFutureUsage{} "I want to use the undo feature in the future"}
Again, we found a significant (\ano{3}{93}{35.27}{<.001}) main effect for the \ivUndoTechnique{} on the \dvDesiredFutureUsage{} rating with a \efETAsquared{0.53} effect size. Post-hoc tests revealed significantly higher ratings for \ivIndividualUndo{} and \ivWorldUndo{} compared to \ivNoUndo{} and \ivSelectiveUndo{} ($p<.001$). We could not find a significant main effect (\ano{1}{31}{3.47}{>.05}) for \ivCollaborationMode{}. Again, we found significant (\ano{3}{93}{4.10}{<.01}) interaction effects with a \efETAsquared{0.11} effect size. While in the \ivCollaborative{} case the \ivWorldUndo{} received non-significant higher ratings compared to \ivIndividualUndo{}, again in the \ivDivided{} case, this inverts, resulting in significantly higher ($p<.05$) ratings for \ivIndividualUndo{} compared to \ivWorldUndo{}. 

These results partially support H2, as only \ivIndividualUndo{} and \ivWorldUndo{} increased the user experience, while \ivSelectiveUndo{} did not.

 \begin{figure*}[t!]
	
    \centering
	\includegraphics[width=\textwidth]{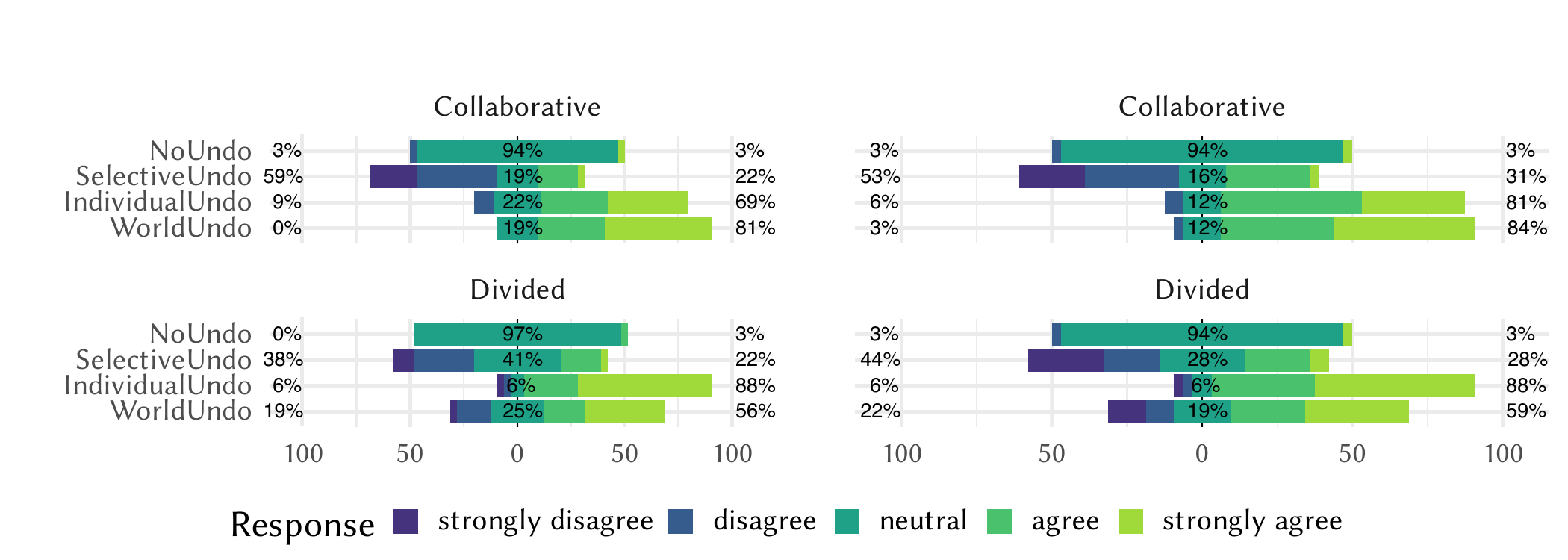}\hfill
	\vspace{-1em}
	\begin{minipage}[t]{.5\linewidth}
		\centering
		\subcaption{\dvEnjoyment}\label{fig:results:Enjoyment}
	\end{minipage}%
    \begin{minipage}[t]{.5\linewidth}
		\centering
		\subcaption{\dvDesiredFutureUsage}\label{fig:results:DesiredFutureUsage}
	\end{minipage}%

	\caption{Participants' responses regarding (a) \dvEnjoyment{} and (b) \dvDesiredFutureUsage{} on a 5-point Likert scale.}
     \Description[Four Likert plots showing the participants responses to "Enjoyment" and "DesiredFutureUsage"]{Four Likert plots showing the participants responses to "Enjoyment" and "DesiredFutureUsage". The two left graphs show the responses for Enjoyment Collaborative (top) and Divided (bottom). The two right graphs show the responses for DesiredFutureUsage Collaborative (top) and Divided (bottom) Each plot consists of four subplots with the responses per technique on a five point Likert scale.}
    \label{fig:results:enjoyment-desired}
\end{figure*}


\subsection{Number of Grab Actions}
We fitted a Poisson regression model to analyze the \dvNumberGrabs{} as an efficiency measure. Here, we found values ranging from $M= 21.4, SD = 8.42$ (\ivCollaborative{}, \ivWorldUndo{}) to $M= 30.8, SD = 11.2$ (\ivDivided{}, \ivNoUndo{}). Our analysis shows a significant (\chisq{1}{27.01}{<.001}) main effect for the \ivCollaborationMode{}. Post-hoc tests revealed significantly ($p<.001$) higher values for \ivDivided{} ($M= 27.7, SD = 10.4$) compared to \ivCollaborative{} ($M= 22.6, SD = 9.19$). We also found a significant (\chisq{3}{21.25}{<.001}) main effect for the \ivUndoTechnique{}. Here, post-hoc tests revealed significantly lower values for \ivWorldUndo{} ($p<.001$)  ($M= 23.4, SD = 9.50$) and \ivIndividualUndo{} ($p<.01$) ($M= 24.2, SD = 9.97$) compared to \ivNoUndo{} ($M= 27.4, SD = 11.3$). We could not find a significant (\chisq{3}{2.29}{>.05}) interaction effect.

These results partially support H1, as only \ivIndividualUndo{} and \ivWorldUndo{} increased the performance, while \ivSelectiveUndo{} did not.



\subsection{Number of Undo Actions}
To analyze the undo actions, we excluded the trials for the \ivNoUndo{} conditions. For the remaining conditions, we fitted a Poisson regression model and
found values ranging from from $M= 12.1, SD = 13.5$ (\ivDivided{}, \ivWorldUndo{}) to $M= 4.69, SD = 7.70$ (\ivCollaborative{}, \ivWorldUndo{}).

We found a  significant (\chisq{1}{23.59}{<.001}) main effect for the \ivCollaborationMode{} on the counted \dvNumberUndos{}. Post-hoc tests revealed significantly ($p<.001$) higher values for \ivDivided{} ($M = 8.42, SD = 11.8$) compared to \ivCollaborative{} ($M = 4.38, SD = 7.10$). We also found a significant (\chisq{2}{6.77}{<.05}) main effect for the \ivUndoTechnique{}. However, post-hoc tests did not confirm this observation ($p>.05$). We found a significant (\chisq{2}{15.80}{<.001}) interaction effect. While the \ivWorldUndo{} received significantly ($p < .05$) lower ratings compared to \ivSelectiveUndo{} in the \ivCollaborative{} case, this inverts in the \ivDivided{} case, where \ivWorldUndo{} received non-significantly ($p > .05$) higher ratings compared to all other levels.

Since these results do not allow for a simple conclusion for H1, H2, and H4, we discuss them in more detail in \autoref{sec:discussion}.

\subsection{Final Tower Height}
As a performance measure, we analyzed the final \dvTowerHeight{}. Performing a two-way RM ANOVA we found values ranging from $M= 4.05, SD = 1.05$ (\ivCollaborative{}, \ivWorldUndo{}) to $M= 2.51, SD = 1.03$ (\ivDivided{}, \ivSelectiveUndo{}). Our RM ANOVA shows a significant (\ano{1}{31}{29.22}{<.001}) main effect for the \ivCollaborationMode{} on the \dvTowerHeight{} with a \efETAsquared{0.11} effect size. Post-hoc tests revealed significantly  ($p<.001$) higher ratings for \ivCollaborative{} ($M= 3.56, SD = 1.15 $) compared to \ivDivided{} ($M = 2.75, SD = 1.13$).
We also found a significant (\ano{2.73}{84.78}{2.99}{<.05}) main effect for the \ivUndoTechnique{} on the \dvTowerHeight{} rating with a \efETAsquared{0.34} effect size. Post-hoc tests revealed significantly  ($p<.05$) higher ratings for \ivWorldUndo{} ($M=3.38, SD= 1.33$) compared to \ivSelectiveUndo{} ($M= 2.82, SD = 1.08)$. We could not find significant (\ano{2.90}{89.83}{2.5}{>.05}) interaction effects.

Again, these results only support H1 and H4 partially, as not all \ivUndoTechnique{}s show a positive effect on the performance and mutual interference. We discuss this in more detail in \autoref{sec:discussion}.

\begin{figure*}[t!]
	\includegraphics[width=\linewidth]{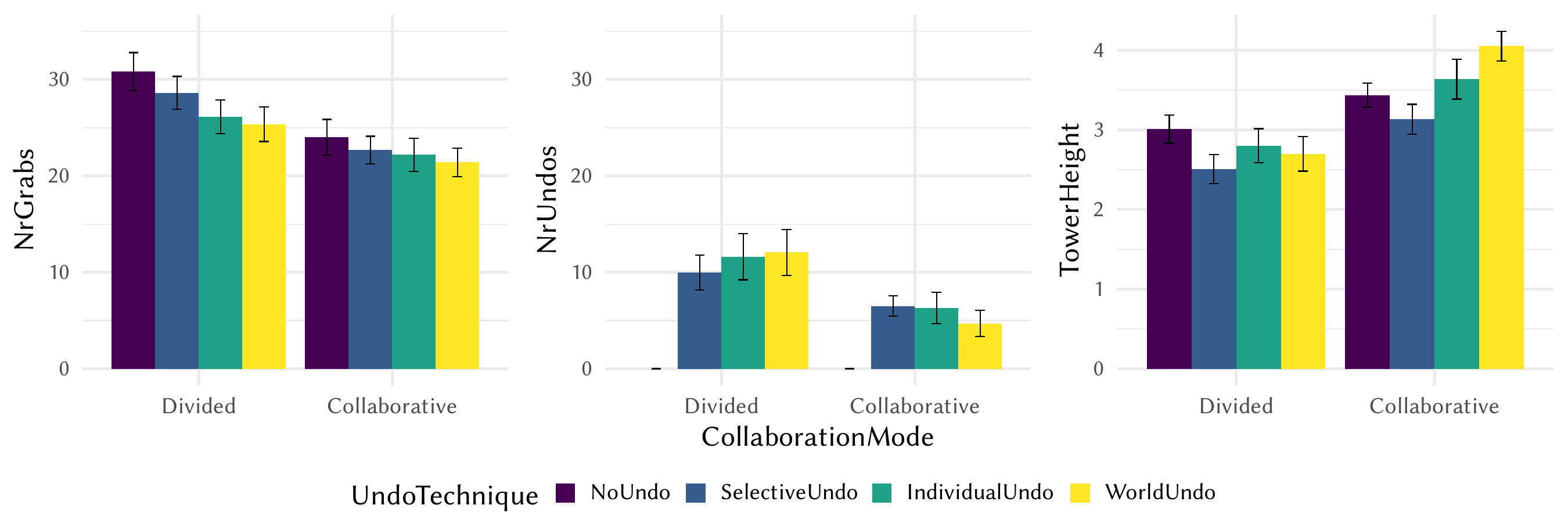}
	\begin{minipage}[t]{.3\linewidth}
		\centering
        \vspace{-1em}
		\subcaption{\dvNumberGrabs{}}\label{fig:results:dvNumberGrabs}
	\end{minipage}%
    \begin{minipage}[t]{.3\linewidth}
		\centering
        \vspace{-1em}
		\subcaption{\dvNumberUndos{}}\label{fig:results:dvNumberUndos}
	\end{minipage}%
     \begin{minipage}[t]{.3\linewidth}
		\centering
        \vspace{-1em}
		\subcaption{\dvTowerHeight{}}\label{fig:results:dvTowerHeight}
	\end{minipage}%

	\caption{The mean results of our Logged Data of the user study. (a) Shows the \dvNumberGrabs{}, (b) the \dvNumberUndos{} on the same scale, and (c) the \dvTowerHeight{}. The error bars show the standard error of the data.}
	\Description[The graphs for the logged data NumberOfGrabs, NumberOfUndos, and TowerHeight]{A figure consisting of three bar charts. Each plot is separated into the Divided and Collaborative section. Each of these section depicts the results for the four undo conditions, NoUndo SelectiveUndo, IndividualUndo, WorldUndo. The first plot shows the NumberOfGrabs count, the second the NumberOfUndos count, and the third the TowerHeight.}
	\label{fig:ShortLogs}
\end{figure*}






\subsection{Qualitative Feedback}

Besides the Questionnaires and Logging, we also collected qualitative feedback from participants through positive and negative comments after each condition. We used a thematic analysis to identify themes within these comments following the process by \citet{blandford_qualitative_2016}. To do so, three researchers individually coded a sample of 15\% of the comments before discussing and agreeing on the final codes together. As a last step, one researcher reviewed the remaining material, coding them with the agreed codes. In this section, we briefly highlight participants' statements using italic typesets for the themes and quotes for participants' statements. In general, participants commented more about the available \ivUndoTechnique{}s than their \ivCollaborationMode{} and their comments aligned with the quantitative results.


When \ivNoUndo{} was available, participants commented on their \emph{worry of mistakes} and bad \emph{recoverability} and stated there was \nquote{no room for errors} (P4) as \nquote{mistakes are devastating} (P9). Participants also \nquote{felt the pressure of not having the undo feature} (P13). As a consequence, participants commented both positively and negatively about their more \emph{thoughtful actions} in these conditions. Another theme within the positive comments was participants' \emph{independence}, and participants liked that they were \nquote{not disturbed by other player} (P3) and \nquote{not disturbing other player} (P11).

For the \ivSelectiveUndo{}, the dominant theme for both positive and negative comments was the \emph{granularity of control}. On the positive side, this was reflected by comments like \nquote{nice to undo one selected object only} (P23) while on the negative side by comments like \nquote{undoing one cube at a time didn't help or made it worse} (P1). Another theme was the need for \emph{memorization}, and participants disliked that they had to \nquote{remember the correct order to undo objects} (P26). This links to the next theme \emph{efficiency}, which was mostly present in the negative comments by statements like \nquote{Undoing something takes a lot of time} (P12). 

After using the \ivIndividualUndo{}, participants positively stated they were \nquote{not afraid to make mistakes} (P9) and had the \nquote{confidence to take risks} (P27), resulting in \emph{no worry of mistakes}. Other themes among the positive comments were the \emph{efficiency} as well as \emph{recoverability} reflected by comments like \nquote{The individual undo works well and lets you recover from big mistakes fast} (P26). 

For \ivWorldUndo{}, participants again positively commented they had \emph{no worry of mistakes} and liked that they \nquote{didn't have to think about my/our mistakes} (P16). While participants commented on this mainly in the \ivCollaborative{} case, in the \ivDivided{} case, the dominant theme among the negative comments was participants \emph{dependency}. This was, on the one hand, expressed by comments regarding the own progress \nquote{I could not accomplish anything since the other player was constantly undermining my efforts with their undo} (P9), but on the other hand, regarding empathy with the other player \nquote{I didn't want to interrupt the other persons flow, so i felt uncomfy in certain situation, especially if I made a mistake and need to use the undo function} (P4).

\section{Discussion}
\label{sec:discussion}
The results presented in \autoref{sec:results} indicate that our participants overall liked and utilized \ivUndoTechnique{}s when available in our study environment. Our qualitative and quantitative data allows insights into participants' usage preference and their acceptance of the different \ivUndoTechnique{}s in different \ivCollaborationMode{}s. In this section we discuss the results with regard to our hypotheses, compare the strengths and shortcomings of the different \ivUndoTechnique{}s, and give recommendations on how to best use them.


\subsection{Undo Increases Perceived Success, but Not Necessarily Actual Success}

Our results indicate that participants used the undo feature whenever available. In particular, the \dvNumberUndos{} proves that participants used all \ivUndoTechnique{}s if they can. However, we observe differences between the techniques.
Independent of the collaboration type, users embraced the \ivIndividualUndo{} and \ivWorldUndo{} as it increased their perceived \dvSuccess{} and helped them recover from mistakes. It also reduced participants' \dvFrustration{} and task load during the task as evident from the \ac{RTLX}-score, supporting H2. 
Further, participants reported that they enjoyed using those techniques and want to use them in the future. 
Contrary to how participants perceived their performance, interestingly, they did not actually perform better throughout by using the available techniques. In the \ivDivided{} case, the \dvTowerHeight{} did not change significantly when utilizing the \ivUndoTechnique{}s. On the other hand, when working \ivCollaborative{}, the usage of \ivWorldUndo{} helped to achieve the highest \dvTowerHeight{}. Consequently, this does not universally support or contradict our H1 as we can observe differences in between the different \ivUndoTechnique{}s and \ivCollaborationMode{}s regarding the actual performance.

We attribute the generally positive participants' response towards the undo features to participants' familiarity with the concept in traditional computer systems and their need to undo mistakes in the context of our study. By providing a mechanism to revert mistakes, we satisfy this need and encourage participants to take more risk in their construction style.
Our results showing divergence of actual and perceived success are in line with findings from previous studies \cite{muller_undoport_2023}. As \citet{archer_user_1984} observe \nquote{The nature of errors may change with experience, but their occurrence does not.} hinting at the fact, that an undo not only aims at increasing performance.
The comments received by participants underline this aspect relevant to undo, namely that they were \nquote{not afraid to make mistakes} and were willing to \nquote{take risks}.

Reflecting on our results, we suggest including undo operations in future \ac{VR} applications to reduce frustration and increase the feeling of success. In \ac{VR} applications for collaborative design processes, such undo mechanics could, for example, motivate users to explore new variations, and users of virtual learning environments could profit from reduced fear of mistakes.






\subsection{World Undo Connects Users, Individual Undo Prevents Mutual Interference}

Besides the task performance, the choice of the \ivUndoTechnique{} also influenced the perceived social connectedness between the participants. This is evident from the ratings on the \ac{IOS} and \ac{GEQ} scores that indicate higher social connectedness for \ivWorldUndo{} compared to \ivIndividualUndo{}, supporting H3. Our data shows that this is especially true for the \ivDivided{} case, where the social connectedness was lower otherwise. While \ivWorldUndo{} increased social connectedness, on the other hand, it also caused an increase in reciprocal disturbances, which backs H4.

We speculate that in the \ivCollaborative{} case, the actual collaboration and common task goal outweigh the influence of the \ivUndoTechnique{}s. For  \ivDivided{} work, however, \ivWorldUndo{} proved to be crucial to fostering social connectedness. We assume that when working separately, the occasional consequence of the other user's use of the \ivWorldUndo{} is sufficient to remind users of their existence and actions. While the \ivWorldUndo{} works as a means to provide awareness of the other user, it can also influence and disturb the other player, as undo actions by one user directly influence the progress of the other user. As expected, this is particularly problematic in the divided case, where users are focused on their own tasks. 







Considering these results, we suggest selecting an undo technique based on weighing the goals and the mode of collaboration between the users. If users work \ivCollaborative{}, the \ivWorldUndo{} does not impose negative effects, and participants favored this technique above all others.
If users work \ivDivided{} and it is relevant not to disturb them, the \ivIndividualUndo{} should be selected, resulting in a lower social connection to other users however.
If users work \ivDivided{} but a high social connection is the design goal, the \ivWorldUndo{} should still be selected, even though this will result in participants' mutual disturbance.



\subsection{Selective Undo Is Not Suitable for Situations Where Large Changes to the Scene Need to Be Reverted}

In contrast to the \ivIndividualUndo{} and \ivWorldUndo{}, the \ivSelectiveUndo{} was not received well for the task at hand. This \ivUndoTechnique{} increased the task load and \dvFrustration{} and helped significantly less in \dvRecover{}ing from mistakes compared to the \ivWorldUndo{} and \ivIndividualUndo{}. When utilizing this technique, participants also felt less \dvSuccess{} and \dvEnjoyment{}.



We attribute these results to the inappropriateness of the chosen form of the \ivSelectiveUndo{} technique for the chosen task, not as a general problem of this type of undo. In the task, undoing only single building blocks was rarely beneficial, as the tower more often collapsed as a connected bigger structure. In these cases our implementation of a \ivSelectiveUndo{} required similar effort by participants as rebuilding a new tower, potentially with better stability. 
Further, the task featured strong dependencies of the manipulated artifacts, and undoing a single mistake did not recover the resulting consequences, like e.g., a more linear undo like \ivIndividualUndo{} or \ivWorldUndo{}.
While related work identified the isolated manipulation and maintaining of subsequent manipulation steps as one strength of the selective undo \cite{berlage_selective_1994}, in our study, this proved not to be useful, as mirrored by participants commenting this technique being \nquote{not helpful}.

This inadequacy of a technique is mirrored in the use of \ivIndividualUndo{} in the \ivCollaborative{} case. The alternating stacking of elements caused a time dependency, limiting users to the last undo step with the \ivIndividualUndo{}.


Following our results, we recommend a thorough assessment of the use cases before selecting an undo technique, in particular, a \ivSelectiveUndo{}. While \ivSelectiveUndo{} was generally rejected in the quantitative and qualitative data in our controlled experiment, we nevertheless consider this \ivSelectiveUndo{} or another possible implementation~\cite{prakash_undoing_1992} suitable for other task types and also advantageous over \ivWorldUndo{} and \ivIndividualUndo{}. Future work is necessary to provide a more in-depth analysis of different implementations of selective undo for \ac{VR}.
\section{Limitations and Future Work}
\label{sec:limitions}
Our data as well as comments from participants prove the need for undo actions in the \ac{VR} domain as well. In the context of our study, we could identify strengths and shortcomings of the various \ivUndoTechnique{}s through questionnaires, qualitative comments, and logged data. As evident from our discussion, this is only a first step in transferring findings from traditional \ac{CSCW} into the \ac{VR} domain and we are confident that this work will serve as a base for future work to further address these challenges. Throughout our study and the consecutive analysis, we identified several limitations imposed by our study design as well as directions for future work, which we discuss in this section.

\subsection{External Validity and Real-World Applicability}

In this paper, we contributed the results of an experiment that explored  \ivUndoTechnique{}s for \ac{VR} in a highly artificial task and environment, while enforcing strict \ivCollaborationMode{}s. As the first work investigating undo mechanics for multi-user systems in \ac{VR}, we adopted such a highly controlled approach to provide a solid foundation for future work. Therefore, we selected representative extremes of collaborative situations on the continuum between joint work on a common goal to independent work with individual goals. Further, we opted for a task that was suitable for both of these collaborative situations and that allowed us to quantify the performance of the participants.
In realistic scenarios, collaboration situations may not fall within these extremes and may be subject to constant change. Further, external influences such as objects that are not part of the actual task or actions that do not result from user interactions may have an impact. We acknowledge that such changes could yield other results and thus further work is needed in this area.

\subsection{Scaleability}
In our experiment, we decided on a dyadic task, as the minimum to study collaborative work and potential usage of \ivUndoTechnique{}s. We expect, that with increasing numbers of users, an \ivUndoTechnique{} like \ivWorldUndo{} can easily become very unfavorable due to the potential for high disturbance, as found already in our dyadic case. At the same time, as discussed before, \ivIndividualUndo{} techniques might be useless in complex scenarios with highly intertwined actions of multiple users. In this context, we see the potential for the exploration of variations of \ivSelectiveUndo{} techniques, balancing between individual understandable yet ineffective and global effective yet chaotic actions.

\subsection{Ownership of Objects in the Shared Virtual Space}
Implementing an \ivIndividualUndo{} allows users to undo their actions. This brings up the question, of what these "own actions" are. In our experiment we only considered the objects manipulated directly by a user, as "their" objects. 
One can ask, if a manipulation resulting from an object manipulated by one player, should also be considered this player's action. The consequent question is which margin of manipulation should one consider here. Which limits for the translation or rotation of an affected object to choose? Could the mere contact with an object be considered manipulation already? With regard to our task, we opted against this implicit definition of manipulation as this implementation would blur the borders of the individual and world undo. However, these are highly relevant questions and future work should address the effect of different implementations of these.


\subsection{Continuous versus Discrete Steps}

During the development of our controlled experiment, we had to make certain design decisions.
In contrast to most undos known from traditional computer systems, we chose to implement continuous undo functions. This obviously is one relevant variable and as shown in related work \cite{muller_undoport_2023} has a strong influence on how users perceive and understand their environment. We chose the continuous implementation, as we derived from related work, that discrete undo steps will impose problems in understanding the mutual undo actions. Future work should investigate these effects of different implementations of a discrete and continuous undo again in the context of multi-user undo. Here, we speculate to find differences concerning performance, and especially mutual understanding of undone actions.
We chose to exclude the time passing between the manipulation of an object and the execution of an undo feature from the undo timeline. In doing so, the users experience an effect right after starting the undo, independent of how long the last action has passed.
We also chose to exclude the other user's avatar from the range of effect of any undo feature, as explained in \autoref{sec:methodology}.
All of these design decisions could be chosen as independent variables. However, studying the influence of all of them was beyond the scope of our work and should be addressed in future work.

\section{Conclusion}
\label{sec:conclusion}
In this paper, we investigated the transferability of established undo techniques from the traditional \ac{CSCW} domain to their use in multi-user collaborative \ac{VR} and derived usage guidelines based on our findings. Our results clearly show that we should provide users with one form of undo in comparable \ac{VR} tasks, as this increases users' perceived success, helps them to recover from mistakes, and reduces participants' frustration and task load. Which form of undo is best suited highly depends on the social constellation as well as the task goal of the \ac{VR} scene. We are confident our work provides essential findings relevant to both follow-up studies as well as designers of \ac{VR} applications. The proposed techniques originating from related work can enrich future collaborative systems by providing users with undo mechanics and therefore improving their experience.
As one key consideration when designing \ivUndoTechnique{}s for multi-user \ac{VR}, one must consider how important social connection between users is and how the mode of collaboration of the users will be. Here, one must weigh between higher social connectedness and potential disturbance between users. 

\bibliographystyle{ACM-Reference-Format}
\bibliography{references}

\end{document}